%% file: session.tex
\documentclass[sigconf]{acmart}

\usepackage{amssymb,amsmath}
\usepackage{booktabs}
\usepackage{graphicx,epsfig}
\usepackage{url}
\usepackage{algorithm,algorithmic}
\usepackage{subfigure}
\usepackage{xcolor,colortbl}

\newcommand{\eqdef}{\overset{def}{=}}

\begin{document}

\copyrightyear{}
\acmYear{}
\setcopyright{acmcopyright}
\acmConference{}{}{}\acmPrice{}\acmDOI{}
\acmISBN{}

\title{An Extended Relevance Model for Session Search}

\author{Nir Levine}
\authornote{Work was done during a summer internship in IBM Research - Haifa.}
\affiliation{%
  \institution{Technion - Israel Institute of Technology}
  \city{Haifa, Israel}
  \postcode{32000}
}
\email{levin.nir1@gmail.com}

\author{Haggai Roitman}
\affiliation{%
  \institution{IBM Research - Haifa}
  \city{Haifa, Israel}
  \postcode{31905}
}
\email{haggai@il.ibm.com}

\author{Doron Cohen}
\affiliation{%
  \institution{IBM Research - Haifa}
  \city{Haifa, Israel}
  \postcode{31905}
}
\email{doronc@il.ibm.com}

\renewcommand{\shortauthors}{Levine et al.}

\begin{abstract}
The session search task aims at best serving the user's information need given her previous search behavior during the session.
We propose an extended relevance model that captures the user's dynamic information need in the session.
Our relevance modelling approach is directly driven by the user's query reformulation (change) decisions and the estimate of how much the user's search behavior affects such decisions.
Overall, we demonstrate that, the proposed approach significantly boosts session search performance.
\end{abstract}

\maketitle


%


\input{intro}
\input{related}

\input{approach}

\input{eval}

\bibliographystyle{plain}\small
\bibliography{bib_min}
\end{document}

%% file: intro.tex
\section{Introduction}
We propose an extended relevance model for session search.
Relevance models aim at identifying terms (words, concepts, etc) that are relevant to a given (user's) information need~\cite{Lavrenko:2001}.
Within a session, user's information need, expressed as a sequence of one or more queries~\cite{Carterette:2016}, may evolve over time. User's search behavior during the session may be utilized as an additional relevance feedback source by the underlying search system~\cite{Carterette:2016}. Given user's session history (i.e., previous queries, result impressions and clicks), the goal of the session search task is to best serve the user's newly submitted query in the session~\cite{Carterette:2016}.

We derive a relevance model that aims at ``tracking" the user's \emph{dynamic information need} by observing the user's search behavior so far during the session. To this end, the proposed relevance model is driven by the user's query reformulation decisions. Our relevance modelling approach relies on previous studies that suggest that user query change decisions may (at least partially) be explained by the previous user search behavior in the session~\cite{Jiang:2016qc,Sloan:2015TM,Yang:2015}.  We utilize the derived relevance model for re-ranking the search results that are retrieved for the current user information need in the session. Overall, we demonstrate that, our relevance modeling approach can significantly boost session search performance compared to many other alternatives that also utilize session data.

%% file: related.tex
\section{Related Work}
 Few previous works have also utilized the session context (i.e., previous queries, retrieved results and clicks) as an implicit feedback source for refining the user's query~\cite{Guan:2014,Shen:2005,Tan:2006,VanGysel:2016}. To this end, the query language model was either combined with the language models of previous queries~\cite{VanGysel:2016} or retrieved (clicked) results~\cite{Shen:2005,Tan:2006}. In addition, different query score aggregation strategies for session search were explored~\cite{Guan:2014}. Yet, none of these previous works have actually considered the user's query change process itself as a possible implicit feedback source.

Several recent works have studied various query reformulation (change) behaviors during search sessions~\cite{Jiang:2016qc,Sloan:2015TM,Yang:2015}.
Among the various features that were studied, \emph{word-level} features were found to best explain the changes in user queries during search sessions~\cite{Jiang:2016qc,Sloan:2015TM}. A notable feature was found to 
be the occurrence of query (changed) words in the contents of results that the user previously viewed or clicked~\cite{Jiang:2016qc,Sloan:2015TM,Yang:2015}. 

Few previous works have also utilized query change for the session search task (e.g.,~\cite{Luo:2015DP,Yang:2015}). Common to such works is the modeling of user queries and their change as states and actions within various Reinforcement Learning inspired query weighting and aggregation schemes~\cite{luo2015designing};
In this work we take a rather more ``traditional" approach, inspired by the relevance model framework~\cite{Lavrenko:2001}.

%% file: approach.tex
\section{Approach}

\subsection{Session model}
Session search is a multi-step process, where at each step $t$, the user may submit a new query $q_{t}$. The search system then retrieves the top-$k$ documents $D_{q_t}^{[k]}$ from a given corpus $\mathfrak{D}$ that best match the user's query\footnote{\small With $k=10$ in the TREC session track benchmarks~\cite{Carterette:2016} that we use later on in our evaluation.}. The user may then examine the results list; each result usually includes a link to the actual content and is accompanied with a summary snippet. The user may also decide to click on one or more of the results in the list in order to examine their actual content. Let $C_{q_t}$ denote the corresponding set of clicked results in $D_{q_t}^{[k]}$. In case the user decides to continue and submits a subsequent query, step $t$ ends and a new step $t+1$ begins.
Let $S_{n-1}$ represent the session history (i.e., user queries, retrieved result documents, and clicked results) that was ``recorded" prior to the \emph{current} (latest) submitted user query $q_{n}$.  On each step 1 $\leq t\leq n-1$, the session history is represented by a tuple $S_{t}=\langle \mathcal{Q}_{t}, \mathcal{D}_{t}, \mathcal{C}_{t}\rangle$. $\mathcal{Q}_{t}=(q_1,q_2,\ldots,q_t)$ is the sequence of queries submitted by the user up to step $t$.
$\mathcal{D}_{t}=(D_{q_{1}}^{[k]},D_{q_{2}}^{[k]},\ldots,D_{q_t}^{[k]})$ is the corresponding sequence of (top-$k$) retrieved result lists. $\mathcal{C}_{t}=(C_{q_{1}},C_{q_{2}},\ldots,C_{q_t})$ further represents the corresponding sequence of user clicks.

\subsection{Information need dynamics}
The session search task is to best answer the current user's query $q_{n}$ while considering $S_{n-1}$~\cite{Carterette:2016}. Let $I$ denote the user's (hidden) information need in the session.
The goal of our relevance modelling approach is, therefore, to better capture the user's information need $I$ which may \emph{evolve} during the session.
In order to capture such dynamics, let $I_{t}$ further represent the user's information need at step $t$. We now assume that, $I_{t}$ depends both on the previous (dynamic) information need $I_{t-1}$ prior to query $q_{t}$ submission and the possible change in such need $\Delta I_{t}\eqdef{I_{t-1}\rightarrow I_{t}}$; $\Delta I_{t}$ is assumed be to implied by the change the user has made to her previous query $q_{t-1}$ to obtain query $q_{t}$.

\subsection{Query change as relevance feedback}
We utilize the user's query reformulation (change) process during the session as an implicit relevance feedback for estimating the change in the user's information need $\Delta I_{t}$. As been suggested by previous works~\cite{Jiang:2016qc,Sloan:2015TM,Yang:2015}, user's query changed terms may actually occur in the contents of previously viewed (clicked) search results in $S_{t-1}$. This, therefore, may (partially) explain how the user decided to reformulate her query from $q_{t-1}$ to $q_{t}$~\cite{Jiang:2016qc,Sloan:2015TM,Yang:2015}.  Our proposed relevance model aims at exploiting such query changed term occurrences within the contents of previously viewed (clicked) results so as to discover those terms $w$ (over some vocabulary $V$) that are the most relevant to the current user's information need $I_{n}$. As a consequence, such terms may be used for \emph{query expansion} aiming to better serve the current user's information need $I_{n}$.

Given query $q_{t}$, compared to the previous query $q_{t-1}$, there can be three main query change types, namely \emph{term retention}, \emph{addition} and \emph{removal}~\cite{Jiang:2016qc,Sloan:2015TM,Yang:2015}.
User term retention, given by the set of terms that appear in both query $q_{t}$ and $q_{t-1}$ and denoted $\Delta{q}_{t}^{\leftrightarrow}$, usually represent the (general) thematic aspects of the user's information need~\cite{Jiang:2016qc,Sloan:2015TM,Yang:2015}. Added terms (denoted $\Delta{q}_{t}^{+}$) are those terms that the user added to query $q_{t-1}$ to obtain query $q_{t}$. A user may add new related terms that were encountered in previous results so as to improve the chance of finding relevant content~\cite{Jiang:2016qc}. On the other hand, a user may remove terms from a previous query $q_{t-1}$ (further denoted $\Delta{q}_{t}^{-}$) in order to terminate a subtask or trying to improve bad performing queries~\cite{Jiang:2016qc}. 


\subsection{Relevance model derivation}

Similar to previous works on relevance models~\cite{Lavrenko:2001}, our goal is to discover those terms $w$ ($\in{V}$) that are the most relevant to the user's information need $I_{n}$; To this end, given the user's current query $q_{n}$ and session history $S_{n-1}$,  let $\theta_{\mathcal{S}_{n}}$ denote our estimate of the relevance (language) model. On each step $1\leq t\leq {n}$, such estimation is given by the following first-order \emph{autoregressive} model:

\vspace{-1mm}
\begin{equation}\label{SRM main}
p(w|\theta_{\mathcal{S}_{t}})\eqdef
                      \gamma_{t}p(w|\theta_{\mathcal{S}_{t-1}})+(1-\gamma_{t})p(w|\theta_{\mathcal{F}_{t}}),
\end{equation}

where $\theta_{\mathcal{F}_{t}}$ now denotes the \emph{feedback model} which depends on the user's (reformulated) query $q_{t}$. While $\theta_{S_{t-1}}$ estimates the dynamic information need prior to step $t$ (i.e., $I_{t-1}$), $\theta_{\mathcal{F}_{t}}$ captures the relative change in such need at step $t$ (i.e., $\Delta I_{t}$). 

$\gamma_{t}$ further controls the relative importance we assign to \emph{model exploitation} (i.e., $\theta_{\mathcal{S}_{t-1}}$) versus \emph{model exploration} (i.e., $\theta_{\mathcal{F}_{t}}$).  $\gamma_{t}$ parameter is dynamically determined based on the relevance model's \emph{self-clarity} at step $t$~\cite{Cronen-Townsend:2002}. Self-clarity estimates how much the prior model $\theta_{\mathcal{S}_{t-1}}$ already ``covers" the feedback model $\theta_{\mathcal{F}_{t}}$; formally:

\vspace{-3mm}
\begin{equation}
\gamma_t\eqdef\gamma\cdot\exp^{-D_{KL}(\theta_{\mathcal{F}_{t}}\parallel\theta_{\mathcal{S}_{t-1}})},
\end{equation}

where $\gamma\in[0,1]$ and $D_{KL}(\theta_{\mathcal{F}_t}\parallel\theta_{\mathcal{S}_{t-1}})$ 
is the \emph{Kullback-Leibler divergence} between the two (un-smoothed) language models~\cite{Zhai:2004}. 
Finally, given $q_{n}$, the current user's query in the session, we derive the relevance model $\theta_{\mathcal{S}_{n}}$ by inductively applying Eq.~\ref{SRM main} (with $\theta_{\mathcal{S}_{0}}\eqdef \textbf{0}$).
We next derive the feedback model $\theta_{\mathcal{F}_t}$.

\subsection{Feedback model derivation}\label{feedback model}
Our estimate of $\theta_{\mathcal{F}_t}$ aims at discovering those terms (in $q_{t}$, $q_{t-1}$ or others in $V$) that are \emph{most relevant to the change} in user's dynamic information need from $I_{t-1}$ to $I_{t}$ (i.e., $\Delta{I_{t}}$).
Given queries $q_{t}$ and $q_{t-1}$, we first classify their occurring terms $w'$ according to their role in the query change. Let $\Delta{q_t}$ further denote the set of terms $w'$ that are classified to the same type of query change (i.e., $\Delta{q_t}\in\{\Delta{q_t}^{\leftrightarrow},\Delta{q_t}^{+},\Delta{q_t}^{-}\}$).

Our relevance model now relies on the fact that query changed terms may also occur within the contents of results that were previously viewed (or clicked) by the user~\cite{Jiang:2016qc,Sloan:2015TM,Yang:2015}. Therefore, on each step $t$, let $F_{t}$ denote the set of results that are used for (implicit) relevance feedback. We determine the set of results to be included in $F_{t}$ as follows. If up to step $t<n$  there is at least one clicked result, then we assign $F_{t}=\bigcup\limits_{1\leq{j}\leq{t}}C_{q_j}$. Otherwise, we first define a \emph{pseudo information need} $\mathcal{Q}_{t}'$. $\mathcal{Q}_{t}'$ represents a (crude) estimate of the user's (dynamic) information need up to step $t$ and is obtained by concatenating the text of all \emph{observed} queries in $\mathcal{Q}_{t}$ (with each query having the same importance, following~\cite{VanGysel:2016}). We then define $F_{t}$ as the set of top-$m$ results in $\bigcup\limits_{1\leq{j}\leq{t}}D_{q_j}$ with the highest \emph{query-likelihood} given $\mathcal{Q}_{t}'$ (representing \emph{pseudo-clicks}). Let $p^{[\mu]}(w|\theta_x)\eqdef\frac{tf(w,x)+\mu\frac{tf(w,\mathfrak{D})}{|\mathfrak{D}|}}{|x|+\mu}$ now denote the \emph{Dirichlet} smoothed language model of text $x$ with parameter $\mu$~\cite{Zhai:2004}.
Inspired by the RM1 relevance model~\cite{Lavrenko:2001}, we estimate $\theta_{\mathcal{F}_t}$ as follows:
\vspace{-3mm}
\begin{equation}\label{eq. feedback model}
p(w|\theta_{\mathcal{F}_t})\eqdef\sum\limits_{d\in{F_t}}p^{[0]}(w|\theta_{d})\cdot\left(\sum\limits_{\Delta{q_t}}p(d|\theta_{\Delta{q_t}})p(\Delta{q_t})\right),
\end{equation}

where $p(\Delta{q_t})$ denotes the (prior) likelihood that, while reformulating query $q_{t-1}$ into $q_{t}$, the user will choose to either add (i.e., $\Delta{q_t}^{+}$), remove (i.e., $\Delta{q_t}^{-}$) or retain (i.e., $\Delta{q_t}^{\leftrightarrow}$) terms. Such likelihood can be pre-estimated~\cite{Sloan:2015TM} (i.e., parameterized); e.g., similarly to the QCM approach~\cite{Yang:2015}. Yet, for simplicity, in this work we assume that every user query change action has the \emph{same odds} (i.e., $p(\Delta{q_t})=\frac{1}{3}$).
Please note that, the main difference between our estimate of $\theta_{\mathcal{F}_t}$ and the RM1 model is in the way the later scores documents in $F_{t}$. Such score in RM1 is based on a given query $q_{t}$~\cite{Lavrenko:2001}, with no further distinction between the role that each query term plays or the fact that some of the terms are actually removed terms that appeared in the previous query $q_{t-1}$. Similar to RM1, we further estimate $p(d|\theta_{\Delta{q_t}})\propto \frac{p(\Delta{q_t}|\theta_{d})}{\sum\limits_{d'\in{F_{t}}}p(\Delta{q_t}|\theta_{d'})}$.

User added or retained terms are those terms that are preferred to be included in the feedback documents $F_t$. On the other hand, removed terms are those terms that should not appear in $F_t$~\cite{Jiang:2016qc}. In accordance, we define:
\begin{equation}
p(\Delta{q_{t}}|\theta_{d})\eqdef\left\{
\begin{array}{lc}
  \prod\limits_{w'\in\Delta{q_{t}}}p^{[\mu]}(w'|\theta_{d}), & \Delta{q_{t}}\in\{\Delta{q}_{t}^{\leftrightarrow},\Delta{q_{t}}^{+}\} \\
  1-\sum\limits_{w'\in\Delta{q}_{t}^{-}}p^{[0]}(w'|\theta_{d}), & \Delta{q_{t}}=\Delta{q_{t}}^{-}
\end{array}
\right.
\end{equation}

In order to avoid \emph{query drift}, on each step $t$, we further anchor the feedback model $\theta_{\mathcal{F}_t}$ to the query model $\theta_{q_{t}}$~\cite{Lavrenko:2001} as follows:
\vspace{-1mm}
\begin{equation}
p(w|\theta_{\mathcal{F}_t}')\eqdef(1-\lambda_{t})p^{[0]}(w|\theta_{q_{t}})+\lambda_{t}p(w|\theta_{\mathcal{F}_t}),
 \end{equation}

where $\lambda_t\eqdef\lambda\cdot sim(q_t,q_n)$ is a \emph{dynamic} query anchoring parameter, $\lambda\in[0,1]$ and $sim(q_t,q_n)$ is calculated using the (idf-boosted) \emph{Generalized-Jaccard} similarity measure; i.e.:
\begin{equation}\small
sim(q_t,q_n)\eqdef\frac{\sum\limits_{w'\in{q_t\cap q_n}}\min\left(tf(w',q_{t}),tf(w',q_{n})\right)\cdot idf(w')}{\sum\limits_{w'\in{q_t\cup q_n}}\max\left(tf(w',q_{t}),tf(w',q_{n})\right)\cdot idf(w')}
\end{equation}

According to $\lambda_t$ definition, the similar query $q_{t}$ is to the current query $q_{n}$, the more relevant is the query change in user's information need $\Delta{I_t}$ (modelled by $\theta_{\mathcal{F}_t}$) is assumed to be to the current user's information need $I_{n}$; Therefore, less query anchoring effect is assumed to be needed using query $q_{t}$.

%% file: eval.tex
\section{Evaluation}
\subsection{Datasets}
\begin{table}[h]
\small
\centering
\label{my-label}
\begin{tabular}{lccc|}
\cline{2-4}
\multicolumn{1}{l|}{\textbf{}}                                   & \multicolumn{1}{c|}{\begin{tabular}[c]{@{}c@{}}\textbf{2011}\\ (train)\end{tabular}}      & \multicolumn{1}{c|}{\begin{tabular}[c]{@{}c@{}}\textbf{2012}\\ (test)\end{tabular}}      & \begin{tabular}[c]{@{}c@{}}\textbf{2013}\\ (test)\end{tabular}      \\ \hline
\multicolumn{1}{|l}{\textbf{Sessions}}   &                                                                                  &                                                                                  &                                                             \\ \hline
\multicolumn{1}{|l|}{Sessions}                                   & \multicolumn{1}{c|}{76}                                                          & \multicolumn{1}{c|}{98}                                                          & 87                                                          \\ \hline
\multicolumn{1}{|l|}{Queries/session}                            & \multicolumn{1}{c|}{3.7$\pm$1.8}                                                 & \multicolumn{1}{c|}{3.0$\pm$1.6}                                                 & 5.1$\pm$3.6                                                 \\ \hline
\multicolumn{1}{|l}{\textbf{Topics}}     &                                                                                  &                                                                                  &                                                             \\ \hline
\multicolumn{1}{|l|}{Sessions/topic}                             & \multicolumn{1}{c|}{1.2$\pm$0.5}                                                 & \multicolumn{1}{c|}{2.0$\pm$1.0}                                                 & 2.2$\pm$1.0                                                 \\ \hline
\multicolumn{1}{|l|}{Judged docs/topic}                          & \multicolumn{1}{c|}{313$\pm$115}                                                 & \multicolumn{1}{c|}{372$\pm$163}                                                 & 268$\pm$117                                                 \\ \hline
\multicolumn{1}{|l}{\textbf{Collection}} &                                                                                  &                                                                                  &                                                             \\ \hline
\multicolumn{1}{|l|}{Name}                                       & \multicolumn{1}{c|}{\begin{tabular}[c]{@{}c@{}}ClueWeb09B\end{tabular}} & \multicolumn{1}{c|}{\begin{tabular}[c]{@{}c@{}}ClueWeb09B\end{tabular}} & \begin{tabular}[c]{@{}c@{}}ClueWeb12B\end{tabular} \\ \hline
\multicolumn{1}{|l|}{\#documents}                                & \multicolumn{1}{c|}{28,810,564}                                                  & \multicolumn{1}{c|}{28,810,564}                                                  & 15,700,650                                                  \\ \hline
\end{tabular}
\caption{\small TREC session track benchmarks}\label{tab:datasets}
\end{table}

Our evaluation is based on the TREC 2011-2013 session tracks~\cite{Carterette:2016} (see benchmarks details in Table~\ref{tab:datasets}). The Category B subsets of the ClueWeb09 (2011-2012 tracks) and
ClueWeb12 (2013 track) collections were used. Each collection has nearly 50M documents. Documents
with spam score below 70 were filtered out. Documents were indexed and searched using the Apache Solr\footnote{\small\url{http://lucene.apache.org/solr/}} search engine. Documents and queries
were processed using Solr's English text analysis (i.e., tokenization, Poter stemming, stopwords, etc).

\begin{table*}[t!]
\small
\centering
\label{my-label}
\begin{tabular}{|l|c|c|c|c|c|c|c|c|}
\hline
                         & \multicolumn{4}{c|}{\textbf{TREC 2012}}                                                                                               & \multicolumn{4}{c|}{\textbf{TREC 2013}}                                                                                                                                                                                        \\ \hline
\textbf{Method} & \textbf{nDCG@10} & \textbf{nDCG}  & \multicolumn{1}{l|}{\textbf{nERR@10}} & \multicolumn{1}{l|}{\textbf{MRR}} & \multicolumn{1}{l|}{\textbf{nDCG@10}} & \multicolumn{1}{l|}{\textbf{nDCG}} & \multicolumn{1}{l|}{\textbf{nERR@10}} & \multicolumn{1}{l|}{\textbf{MRR}} \\ \hline
Initial retrieval      & 0.249$^{rq}$            & 0.256$^{rq}$          & 0.302$^{rq}$                                                         & 0.594$^{rq}$                                                     & 0.113$^{rq}$                                                         & 0.105$^{rq}$                                                      & 0.140$^{rq}$                                                         & 0.390$^{rq}$                                                     \\ \hline
FixInt~\cite{Shen:2005}          & 0.333$^{rq}$            & 0.296$^{rq}$          & 0.380$^{rq}$                                                         & 0.679$^{rq}$                                                     & 0.165$^{rq}$                                                         & 0.132$^{rq}$                                                      & 0.209$^{rq}$                                                         & 0.544$^{rq}$                                                     \\ \hline
BayesInt~\cite{Shen:2005}        & 0.334$^{rq}$            & 0.297$^{rq}$          & 0.382$^{rq}$                                                         & 0.674$^{rq}$                                                     & 0.171$^{rq}$                                                         & 0.131$^{rq}$                                                      & 0.208$^{rq}$                                                         & 0.527$^{rq}$                                                     \\ \hline
BatchUp~\cite{Shen:2005}         & 0.320$^{rq}$            & 0.288$^{rq}$          & 0.368$^{rq}$                                                         & 0.664$^{rq}$                                                     & 0.181$^{rq}$                                                         & 0.134                                                      & 0.233$^{rq}$                                                         & 0.581$^{rq}$                                                     \\ \hline
LongTEM~\cite{Tan:2006}        & 0.332$^{rq}$            & 0.295$^{rq}$          & 0.389$^{rq}$                                                         & 0.667$^{rq}$                                                     & 0.167$^{rq}$                                                         & 0.131$^{rq}$                                                      & 0.205$^{rq}$                                                         & 0.530$^{rq}$                                                     \\ \hline
RM3($q_{n}$)~\cite{Lavrenko:2001} & 0.311$^{rq}$            & 0.284$^{rq}$          & 0.369$^{rq}$                                                         & 0.654$^{rq}$                                                     & 0.134$^{rq}$                                                         & 0.122$^{rq}$                                                      & 0.161$^{rq}$                                                         & 0.422$^{rq}$                                                     \\ \hline
RM3($\mathcal{Q}_{n}'$)        & 0.305$^{rq}$            & 0.284$^{rq}$          & 0.354$^{rq}$                                                         & 0.647$^{rq}$                                                     & 0.153$^{rq}$                                                         & 0.129$^{rq}$                                                      & 0.203$^{rq}$                                                         & 0.553$^{rq}$                                                     \\ \hline
QA(uniform)~\cite{VanGysel:2016}     & 0.301$^{rq}$            & 0.282$^{rq}$          & 0.352$^{rq}$                                                         & 0.646$^{rq}$                                                     & 0.160$^{rq}$                                                         & 0.130$^{rq}$                                                      & 0.204$^{rq}$                                                         & 0.546$^{rq}$                                                     \\ \hline
QA(decay)~\cite{Guan:2014}       & 0.303$^{rq}$            & 0.284$^{rq}$          & 0.353$^{rq}$                                                         & 0.645$^{rq}$                                                     & 0.163$^{rq}$                                                         & 0.131$^{rq}$                                                      & 0.207$^{rq}$                                                         & 0.550$^{rq}$                                                     \\ \hline
QCM~\cite{Yang:2015}        & 0.329$^{rq}$            & 0.262$^{rq}$          & 0.306$^{rq}$                                                         & 0.574$^{rq}$                                                     & 0.158$^{rq}$                                                         & 0.129$^{rq}$                                                      & 0.201$^{rq}$                                                         & 0.535$^{rq}$                                                     \\ \hline
QCM(SAT)~\cite{Yang:2015}        & 0.298$^{rq}$            & 0.281$^{rq}$          & 0.347$^{rq}$                                                         & 0.635$^{rq}$                                                     & 0.158$^{rq}$                                                        & 0.129$^{rq}$                                                      & 0.202$^{rq}$                                                         & 0.545$^{rq}$                                                     \\ \hline
QCM(DUP)~\cite{Yang:2015}        & 0.299$^{rq}$            & 0.281$^{rq}$          & 0.350$^{rq}$                                                         & 0.631$^{rq}$                                                     & 0.160$^{rq}$                                                         & 0.130$^{rq}$                                                      & 0.208$^{rq}$                                                         & 0.559$^{rq}$                                                     \\ \hline
\textbf{SRM(RM1)}             & 0.348$^q$  & 0.300 & 0.395$^q$                                                & 0.699$^q$                                            & 0.188$^q$                                                & 0.137                                             & 0.240$^q$                       &                         0.601$^q$                                            \\ \hline
\textbf{SRM(QC)}             & \textbf{0.356}$^{r}$   & \textbf{0.304} & \textbf{0.405}$^{r}$                                                & \textbf{0.716}$^{r}$                                            & \textbf{0.193}$^{r}$                                                & \textbf{0.138}                                             & \textbf{0.248}$^{r}$                                                & \textbf{0.612}$^{r}$                                            \\ \hline
\end{tabular}
\caption{\small Evaluation results. The $r$ and $q$ superscripts denote significant difference with \textbf{SRM(RM1)} and \textbf{SRM(QC)}, respectively ($p<0.05$).} 
\label{tab:results}
\end{table*}

\subsection{Baselines}
We compared our proposed relevance modelling approach (hereinafter denoted \textbf{SRM}\footnote{\small Stands for ``\emph{Session-Relevance Model}".}) with several different types of baselines.
This includes state-of-the-art language modeling methods that utilize session context data (i.e., previous queries, viewed or clicked results); namely \textbf{FixedInt}~\cite{Shen:2005}  (with $\alpha=0.1$, $\beta=1.0$ following~\cite{Shen:2005}) and its Bayesian extension \textbf{BayesInt}~\cite{Shen:2005} (with $\mu=0.2$, $\nu=5.0$, following~\cite{Shen:2005}) -- both methods combine the query $q_{n}$ model with the history queries $\mathcal{Q}_{n-1}$ and clicks $\mathcal{C}_{n-1}$ centroid models; \textbf{BatchUp}~\cite{Shen:2005} (with $\mu=2.0$, $\nu=15.0$, following~\cite{Shen:2005}) which iteratively interpolates the language model of clicks that occur up to each step $t$ using a batched approach; and the \emph{Expectation Maximization} (EM) based approach~\cite{Tan:2006} (hereinafter denoted \textbf{LongTEM} with $\lambda_{q}=0$, $\sigma_{C}=20$ and $\sigma_{NC}=1$, following~\cite{Tan:2006}), which first interpolates each query $q_{t}$ model with its corresponding session history model (based on both clicked (C) and non-clicked (NC) results in $D^{[k]}_{q_{t}}$); the (locally) interpolated query models are then combined based on their relevant session history using the EM-algorithm~\cite{Tan:2006}.

Next, we implemented two versions of the Relevance Model \cite{Lavrenko:2001}. The first is the basic RM3 model, denoted \textbf{RM3}($q_{n}$), learned using the last query $q_{n}$ and the top-$m$ retrieved documents as pseudo relevance feedback. The second, denoted \textbf{RM3}($\mathcal{Q}_{n}'$), uses the pseudo information need $\mathcal{Q}_{n}'$ (see Section~\ref{feedback model}) instead of $q_{n}$ .
We also implemented two query aggregation methods, namely: \textbf{QA(uniform)} which is equivalent to submitting $\mathcal{Q}_{n}'$ as the query~\cite{VanGysel:2016}; the second, denoted \textbf{QA(decay)}, further applies an exponential decay approach to prefer recent queries to earlier ones (with decay parameter $\gamma=0.92$, following~\cite{Guan:2014,Yang:2015}). 
We further implemented three versions of the \emph{Query Change Model} (QCM) -- an MDP-inspired query weighting and aggregation approach~\cite{Yang:2015}. Following~\cite{Yang:2015} recommendation, QCM's parameters were set as follows $\alpha=2.2$, $\beta=1.8$, $\epsilon=0.07$, $\delta=0.4$ and $\gamma=0.92$. The three QCM versions are the basic \textbf{QCM} approach~\cite{Yang:2015}; \textbf{QCM(SAT)} which utilizes only ``satisfied" clicks (i.e., clicks whose dwell-time is at least 30 seconds~\cite{Yang:2015}); and \textbf{QCM(DUP)} which ignores duplicate session queries~\cite{Yang:2015}.

Finally, in order to evaluate the relative effect of the query-change driven feedback model (i.e., $\theta_{\mathcal{F}_t}$), we implemented a variant of SRM by replacing the query-change driven score of Eq.~\ref{eq. feedback model} with the RM1 document score (i.e., $p(d|q_{n})$). Let  \textbf{SRM(QC)} and \textbf{SRM(RM1)} further denote the query-change and ``RM1-flavoured" variants of \textbf{SRM}, respectively. It is important to note that, \textbf{SRM(RM1)} still relies on the dynamic relevance model updating formula (see Eq.~\ref{SRM main}) and the dynamic coefficients $\gamma_{t}$ and $\lambda_{t}$ -- both further depend on the session dynamics (captured by $\theta_{S_{t-1}}$ and $\theta_{\mathcal{F}_t}$).

\subsection{Setup}
Our evaluation is equivalent to the TREC 2011-2012 RL4 and TREC 2013 RL2 sub-tasks~\cite{Carterette:2016}. To this end, given each session's (last) query $q_{n}$, we first retrieved the top-2000 documents with the highest query likelihood (QL) score\footnote{\small For this we used Solr's \texttt{LMSimilarity} with Dirchlet smoothing parameter $\mu=2500$ which is similar to Indri's default parameter.} to $q_{n}$. Documents were then re-ranked using the various baselines by multiplying their (initial) QL score with the score determined by each method. The document scores of the various language model baselines (i.e., \textbf{FixInt}, \textbf{BayesInt}, \textbf{BatchUp}, \textbf{LongTEM} and the variants of \textbf{RM3} and \textbf{SRM} ) were further determined using the KL-divergence score~\cite{Zhai:2004}; where each baseline's learned model was clipped using a fixed cutoff of 100 terms~\cite{Zhai:2004}. The TREC session track \texttt{trec\_eval} tool\footnote{\small\url{http://trec.nist.gov/data/session/12/session_eval_main.py}} was used for measuring retrieval performance. Using this tool, we measured the nDCG@10, nDCG (@2000), nERR@10 and MRR of each baseline.
Finally, we tuned the \textbf{RM3} and \textbf{SRM}'s free parameters\footnote{\small$\lambda\in\{0.1,0.2,\ldots,0.9\}$, $\gamma\in\{0.1,0.2,\ldots,0.9\}$, $m\in\{5,10,\ldots,100\}$} using the TREC 2011 track as a train set. The parameters were optimized so as to maximize MAP. The TREC 2012-2013 tracks were used as the test sets. 

\pagebreak

\subsection{Results}
The evaluation results are summarized in Table~\ref{tab:results}. The first row reports the quality of the \emph{initial} retrieval. Overall, compared to the various alternative baselines, the two \textbf{SRM} variants provided significantly better performance; with \textbf{at least} $+6.6\%$, $+2.4\%$, $+4.1\%$ and $+5.3\%$ better performance in nDCG@10, nDCG, nERR@10 and MRR, respectively, for both test benchmarks. The results clearly demonstrate the dominance of the session-context sensitive language modeling approaches (and the two \textbf{SRM} variants among them) over the other alternatives we evaluated. Furthermore, \textbf{SRM}'s consideration of the user's query-change process as an additional relevance feedback source results in a more accurate estimate of the user's information need. 

Next, compared to the \textbf{RM3} variants, it is clear from the results that a dynamic relevance modeling approach that is driven by query-change (such as \textbf{SRM}) is a better choice for the session search task. Moving from an \emph{ad-hoc} relevance modelling approach (i.e., one that only focuses on the last query in the session) to a session-context sensitive approach provides significant boost in performance; with \textbf{at least} $+14\%$, $+7.0\%$, $+9.8\%$ and $+9.5\%$ improvement in nDCG@10, nDCG, nERR@10 and MRR, respectively, for both test benchmarks.

We further observe that, compared to the baseline methods that implement various query aggregation and scoring schemes (i.e., \textbf{QA} and \textbf{QCM} variants), a query-expansion strategy based on the user's dynamic information need (such as the one implemented by \textbf{SRM} variants) provides a much better alternative; with \textbf{at least} $+18.5\%$, $+6.1\%$, $+15.1\%$ and $+9.5\%$ improvement in nDCG@10, nDCG, nERR@10 and MRR, respectively, for both test benchmarks.

Finally, comparing the two \textbf{SRM} variants side-by-side, it becomes even more clear that, using the query-change as an additional relevance feedback source is the better choice; with \textbf{at least} $+2.3\%$, $+1.0\%$, $+2.5\%$ and $+1.8\%$ improvement in nDCG@10, nDCG, nERR@10 and MRR, respectively, for both test benchmarks. Please recall that, \textbf{SRM(QC)} was trained with a fixed and equal-valued priors $p(\Delta q_{t})$. Hence, a further improvement may be obtained by better tuning of these priors. 